\documentclass[prl,aps,twocolumn,showpacs]{revtex4}
\usepackage{epsfig}
\begin{document}
\title{Quantum computation using the Aharonov-Casher set up}
\author{Marie Ericsson and Erik Sj\"oqvist}
\affiliation{Department of Quantum Chemistry,
Uppsala University, Box 518, S-751 20 Sweden}
\begin{abstract}
It is argued that the Aharonov-Casher set up could be used
as the basic building block for quantum computation. We
demonstrate explicitly in this scenario one- and two-qubit
phase shift gates that are fault tolerant to deformations of
the path when encircling two sites of the computational
system around each other.
\end{abstract}
\pacs{03.65.Vf, 03.67.Lx}
\maketitle

Efficient and reliable quantum computation of arbitrarily long
duration is possible even with faulty components, if the errors
can be corrected faster than they occur \cite{shor95,steane95}.
Even more desirable would be to implement quantum computation
that is intrinsically fault tolerant as this would prevent
errors to occur. Topological ideas arise naturally in this
context, as they are robust against deformations of the
physical path, such as what happens when a small amount of noise
is added.

Motivated by this intuition, fault tolerant quantum
computation based upon topological phases in various anyonic
systems has been suggested recently
\cite{kitaev97,ogburn99,lloyd00,freedman01,mochon02}. Related to
these ideas is to achieve fault tolerance by using the geometric
phase \cite{zanardi99,pachos00,duan01,jones00,ekert00,xiang-bin01},
as this phase is dependent only upon the area enclosed in the
physical state space and is thereby robust against area preserving
deformations of the physical path. These topological and
geometric approaches to fault tolerant quantum computation
have been shown \cite{zanardi02} to be particular cases of
the notion of computation in noiseless quantum subsystems
\cite{knill00}.

In this Letter, we propose the two dimensional Aharonov-Casher
(AC) set up \cite{aharonov84} as the basic building block for
topological one- and two-qubit phase shift gates. This
implementation would be fault tolerant as the AC effect
is only dependent upon the winding number of the physical
path. We suggest that combining the present phase shift gates
with appropriate nontopological one-qubit gates would be a
realisation of universal quantum computation
\cite{lloyd95,deutsch95} based upon the AC set up.

Let us first briefly review the AC effect. In the
two dimensional AC set up, a magnetic moment ${\bf \mu}$
and a point charge $q$ are free to move in the $x-y$ plane,
say. This system is described by the Galilean and gauge
invariant Lagrangian
\begin{equation}
{\cal L} = \frac{1}{2} mv^2 + \frac{1}{2} MV^2 + q{\bf A}
({\bf r} - {\bf R}) \cdot [{\bf v} - {\bf V}] ,
\end{equation}
where $m,{\bf r},{\bf v}$ denote the mass, position, and
velocity, respectively, of the charge, and $M,{\bf R},{\bf V}$
the corresponding quantities of the dipole. In this two dimensional
set up, the curl $\partial_{x} A_{y} - \partial_{y} A_{x}$ of
the vector potential ${\bf A} = (A_{x},A_{y})$ vanishes
except at the origin, but when encircling the two particles
$n$ times around each other along the path $C$ there is a
phase of the form
\begin{equation}
\gamma = \frac{q}{\hbar} \oint_{C} {\bf A} ({\bf l}) \cdot d{\bf l}
\propto n \mu q ,
\end{equation}
where ${\bf l} = {\bf r} - {\bf R}$. This phase is topological
in the sense that it only depends upon the winding number $n$,
which makes it insensitive to small deformations of the path
$C$.

\begin{figure}[h]
\begin{center}
\includegraphics[width=8 cm]{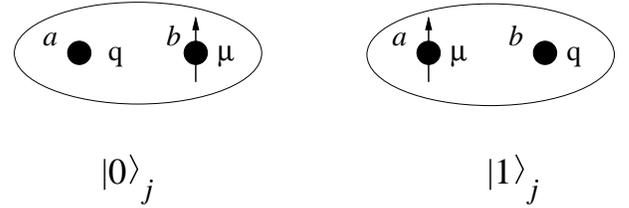}
\end{center}
\caption{The computational basis $|0\rangle_{j}$ and
$|1\rangle_{j}$, stored on different spatial Aharonov-Casher
configurations.}
\end{figure}

This AC scenario may be used to implement fault tolerant one- and
two-qubit phase shift gates as follows. Let us store the
$j^{\text{th}}$ qubit on the two spatial sites $a$ and $b$ of a
single AC set up. The computational basis consists of
$|0\rangle_{j} = |q (a) \mu (b) \rangle_{j}$ with the magnetic
moment localised at site $a$ and the charge localised at site $b$,
respectively, and the reverse for the orthogonal state, i.e.
$|1\rangle_{j} = |\mu (a) q (b) \rangle_{j}$, see Fig. 1. We
assume that the states are sufficiently localised so that $\langle
0|1 \rangle_{j} =0$ for each $j$. The computational system is
build up from such AC-qubits arranged along a line in the two
dimensional plane and let us assume for simplicity that all the
qubits contain the same magnetic moment $\mu$ and the same charge
$q$.

A controlled phase shift gate $B (\gamma)$ can be achieved by
encircling along the path $C$ the particle at site $b$ around
the particle at site $a$ from two different AC-qubits $j$ and
$j'$. This results in
\begin{eqnarray}
B(\gamma): |00 \rangle_{jj'} & \longrightarrow &
e^{i\gamma} |00 \rangle_{jj'}
\nonumber \\
B(\gamma): |01 \rangle_{jj'} & \longrightarrow &
|01 \rangle_{jj'}
\nonumber \\
B(\gamma): |10 \rangle_{jj'} & \longrightarrow &
|10 \rangle_{jj'}
\nonumber \\
B(\gamma): |11 \rangle_{jj'} & \longrightarrow &
e^{i\gamma} |11 \rangle_{jj'} ,
\label{eq:cpsg}
\end{eqnarray}
as illustrated in Fig. 2. This gate is topological as it is
insensitive to any deformations of the path $C$ under the assumption
that charge-charge and dipole-dipole interactions can be neglected
between all pairs of AC-qubits. Except for the trivial case
where $\gamma$ is an integer multiple of $\pi$, $B(\gamma)$
may entangle the qubits it acts on.

Universal quantum computation can be achieved by combining
$B(\gamma)$ with the one-qubit logic gates
\begin{eqnarray}
U (\gamma) & = &
\exp \big[ -i\frac{\gamma}{2} \sigma_{z}^{j}
\big] , \ \ {\text{(phase shift gate)}} ,
\nonumber \\
U_{SWAP} (\theta_j) & = &
\exp \big[ -i\frac{\theta_{j}}{2} \sigma_{y}^{j}
\big] , \ \ {\text{(partial swap gate)}} ,
\nonumber \\
\end{eqnarray}
where $\sigma_{z}^{j} = |0\rangle_{j} \langle 0|_{j} -
|1\rangle_{j} \langle 1|_{j}$ and $\sigma_{y}^{j} =
-i|0\rangle_{j} \langle 1|_{j} + i|1\rangle_{j} \langle 0|_{j}$.
That is, any $N-$qubit logic operation can be simulated to any
precision with an appropriate set of $U (\gamma),U_{SWAP} (\theta_j)$,
and $B (\gamma)$ gates. The one-qubit phase shift gate $U (\gamma)$
is achieved by encircling the two sites within a single AC-qubit
around each other depending upon whether the state is $|0\rangle_j$
or $|1\rangle_j$. This could for example be achieved by addressing
the site $a$, say,
in such a way that the particle there is only taken around site
$b$ if it is charged. This would result in $U(\gamma)$ up to
an unimportant overall phase factor $e^{i\gamma /2}$.  The
$U_{SWAP} (\theta_j)$ gates could be realised in principle by
exposing beam-splitters to each of the AC set ups. The parameter
$\theta_j$ determines the transmission probability $T$ of such
a beam-splitter according to $T=\cos (\theta_j /2)$.

\begin{figure}[h]
\begin{center}
\includegraphics[width=8 cm]{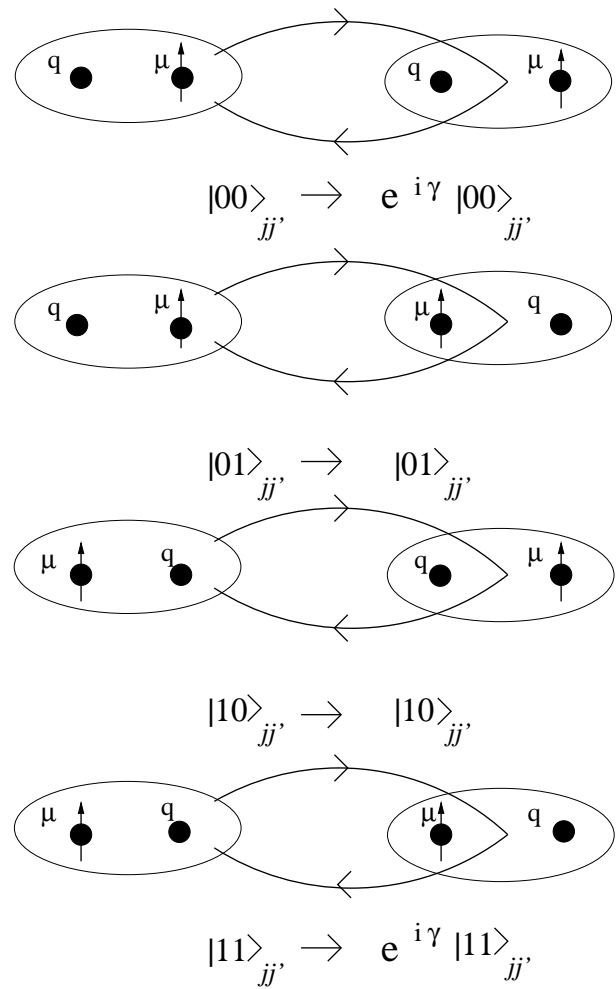}
\end{center}
\caption{Controlled phase shift gate based upon the
Aharonov-Casher set up.}
\end{figure}

While the $U (\gamma)$ gate is topological and thereby fault
tolerant to path deformations, the $U_{SWAP} (\theta_j)$ gate is
essentially dynamical and nontopological as it relies upon
the detailed interaction between the AC-qubit and the beam-splitter.
This situation is expected since the present treatment of the AC
effect is basically Abelian and there are therefore non-Abelian
operations necessary to achieve universality that could not be
obtained by the AC effect alone (see Ref. \cite{lloyd00} for a
similar case).

Quantum computation based upon the AC set up can be realised
as follows. First, translate the quantum algorithm into a set
of elementary one- and two-qubit gates. Prepare an initial
state by spreading out a set of AC composites along a line
in the plane of motion. Perform appropriate phase shifts and
partial swaps on each AC set up and perform appropriate
two-qubit controlled phase shifts by braiding sites from
pairs of AC-qubits. The final answer of the computation
is obtained by measuring the spatial location of the
particles in the output.

Universal quantum computation using the AC set up only
involves electromagnetic interactions between elementary
systems and works even for distinguishable qubits. This
should be compared with the suggestions for topological
quantum computation in Refs.
\cite{kitaev97,ogburn99,lloyd00,freedman01,mochon02},
that all involves collective effects such as anyons or
spin systems with long-range correlations.

In principle, quantum computation based upon the AC set up could be
realisable in combined atom-ion systems confined to a plane. A similar
implementation could be achieved in three dimensions by replacing the
point charge with a line of charge.  However, to put this charged line
in a coherent superposition would be difficult in practise and it is
therefore unclear whether AC based quantum computation could have any
relevance in the three-dimensional context. Moreover, it is
important to keep in mind that the AC shift is essentially a
relativistic effect and thus usually quite small also for such systems
(see, e.g., \cite{sangster93,gorlitz95}), which means that the gates
have to be repeated many times to achieve phase shifts of useful
size. This may spoil the fault tolerance of the phase shift gates as
the error probability is expected to increase with the winding
number. Another challenge, associated with the implementation of the
controlled phase shift gate, is the control of the nontopological
charge-charge and dipole-dipole interactions that act between the
various AC-qubits. These interactions could be made small under
certain circumstances, but may add up when repeating the gate.

In conclusion, we have proposed to use the two dimensional
Aharonov-Casher (AC) set up as the basic building block for
quantum computation. We have argued that the AC set up could
be useful in the implementation of one- and two-qubit phase
shift gates that are fault tolerant to path deformations
when two sites are encircled around each other. Universality
is achieved by adding nontopological one-qubit partial swap
gates. Although it seems hard to implement quantum computation
based upon the AC set up with present day technology, we believe
it has a conceptual value as it demonstrates topological quantum
computation using electromagnetic interactions between elementary
systems.
\vskip 0.5 cm
The work by E.S. was financed by the Swedish Research Council.

\end{document}